%% file: main.tex
\documentclass[letterpaper]{article} 
\usepackage[draft]{aaai2026}  
\usepackage{times}  
\usepackage{helvet}  
\usepackage{courier}  
\usepackage[hyphens]{url}  
\usepackage{graphicx} 
\usepackage{natbib}  
\usepackage{caption} 
\usepackage{array}
\usepackage{booktabs,tabularx,ragged2e}
\newcolumntype{Y}{>{\RaggedRight\arraybackslash}X}
\usepackage{adjustbox}
\usepackage{algorithmic}
\usepackage{booktabs}
\usepackage{multirow}
\usepackage{siunitx}
\usepackage{makecell}
\usepackage{amsmath}
\usepackage{tabularx}
\usepackage{enumitem}
\usepackage{framed}
\usepackage{amssymb}
\usepackage[ruled,linesnumbered]{algorithm2e}
\usepackage{bm}
\usepackage{amsfonts}       
\usepackage{nicefrac}       
\usepackage{microtype}      
\usepackage[dvipsnames]{xcolor}
\usepackage{listings}
\usepackage{subcaption}
\usepackage{url}
\usepackage{dblfloatfix} 
\usepackage{placeins}
\usepackage{soul}

\title{News-Aware Direct Reinforcement Trading for Financial Markets}
\author{
  Qing-Yu Lan, Zhan-He Wang, Jun-Qian Jiang, \\ Yu-Tong Wang, and Yun-Song Piao\\
}
\affiliations{
    School of Physical Sciences\\
  University of Chinese Academy of Sciences\\
  Beijing, 100049, China \\
   \{lanqingyu19, wangzhanhe19, jiangjunqian21\}@mails.ucas.ac.cn,\\  \{wangyutong, yspiao\}@ucas.ac.cn\\
}

\DontPrintSemicolon
\SetAlgoLined

\SetCommentSty{mycommentsty}
\SetKwComment{Comment}{\color{RoyalBlue}\footnotesize$\triangleright$\ }{}

\begin{document}

\maketitle

\begin{abstract}
\input{sections/Abstract}
\end{abstract}

\input{sections/Introduction}

\input{sections/Related_Work}

\input{sections/Methodology_v3}

\input{sections/Experiment}

\input{sections/Conclusion}


\clearpage
\appendix

\setcounter{secnumdepth}{1}
\input{sections/Appendix}

\end{document}

%% file: sections/Abstract.tex
The financial market is known to be highly sensitive to news.
Therefore, effectively incorporating news data into quantitative trading remains an important challenge. 
Existing approaches typically rely on manually designed rules and/or handcrafted features.
In this work, we directly use the news sentiment scores derived from large language models, together with raw price and volume data, as observable inputs for reinforcement learning.
These inputs are processed by sequence models such as recurrent neural networks or Transformers to make end-to-end trading decisions.
We conduct experiments using the cryptocurrency market as an example and evaluate two representative reinforcement learning algorithms, namely Double Deep Q-Network (DDQN) and Group Relative Policy Optimization (GRPO).
The results demonstrate that our news-aware approach, which does not depend on handcrafted features or manually designed rules, can achieve performance superior to market benchmarks.
We further highlight the critical role of time-series information in this process.

%% file: sections/Introduction.tex
\section{Introduction}
\label{sec:introduction}

The inherent complexity and volatility of financial markets pose significant challenges to high-quality investment decision-making and undermine the reliability of traditional trading signals \cite{Hambly2023Recent}, especially for cryptocurrency markets \cite{Dro2023what,Wei2023Crypto}.
In tasks such as stock portfolio management, each decision is usually driven by an integrated and diverse information flow with varying timeliness and forms, including market data, technical indicators, and market sentiment. Manual trading struggles to process these signals at scale and stay consistent under time pressure, which slows execution and weakens risk control. This motivates automated, data-driven quantitative systems that can fuse diverse signals and optimize returns while managing current market risk.  

Reinforcement Learning (RL) serves as a powerful framework for building automated quantitative trading systems, enabling agents to explore complex market environments and continually update their policies to optimize strategies and maximize returns.
However, markets are driven not only by past prices and volumes but also by news, which delivers extra information shocks and triggers market regime changes. Leaving out news makes the whole market condition only partially observable and increases non-stationarity. In early studies, financial news was processed with dictionary-based methods (e.g.  sentiment lexicons), the appearance of large language models (LLMs) (e.g.~\cite{Brown2020Language}) offers the possibility of automated unstructured textual information processing, rapidly extracting sentiment signals
from news text,
thereby enabling the development of RL agents that leverage news datasets in a reliable manner. 
Therefore, in recent years, LLMs have been investigated for financial trading and portfolio management, with strong results in sentiment extraction and explanation generation.
\begin{figure*}[t!]
  \vspace{-\intextsep}
  \centering
  \includegraphics[width=0.99\linewidth]{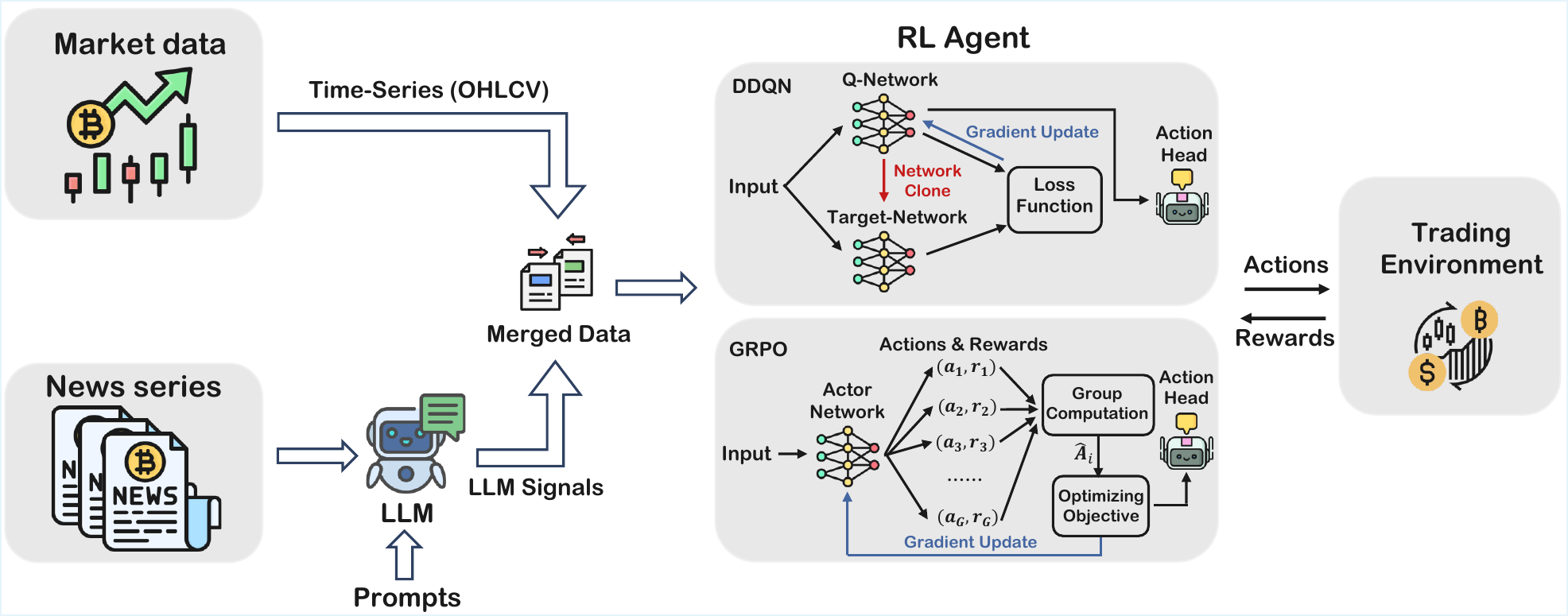}
  \caption{Schematic of the proposed news-aware reinforcement learning framework for financial trading. The system comprises an LLM that analyzes financial news; its output signals are combined with market data and then input to an RL agent. The RL agent utilizes DDQN/GRPO algorithms with various network architectures to process the integrated time-series data, and the action head's output actions interact with the trading environment to generate rewards.
}
  \label{fig:pipeline}
  \vspace{-\intextsep}
\end{figure*}

Despite the success of RL methods in data-driven decision making, in financial markets, successful RL approaches typically rely on well-designed handcrafted features. These technical factors or indicators were developed for equity markets and may not be applicable to the cryptocurrency markets. Moreover, handcrafted technical indicators often generalize poorly. For instance, the moving average feature can effectively capture trends but may incur substantial losses in mean-reversion markets \cite{poterba1988mean}.
Similarly, the utilization of news information is also based on manually designed rules.

Based on these concerns, in this paper, we explore whether news data can be incorporated directly with raw price and volume as observable inputs to the RL agent without handcrafted features or manually designed rules.
The overview of our pipeline is shown in Figure~\ref{fig:pipeline}.
We extract sentiment from finance-related news using an LLM, convert it into structured features (e.g. sentiment scores and risk scores)
and then integrate them with raw market prices and volume.
Within the RL agent, an LSTM or Transformer encoder is employed as the front-end network to process the merged time-series inputs.
We tune hyperparameters on the validation set and conduct backtesting on the test set, finding that the framework can achieve competitive performance relative to market benchmarks and to agents without LLM-derived news sentiment or sequence-based network.

The remainder of this paper is organized as follows. Section~\ref{related work} reviews related work. Section~\ref{methodology} details our methodology, including how we use LLM to conduct sentiment features extraction, the reinforcement learning architectures, and the procedures for dataset preprocessing and hyper-parameters tuning. Section~\ref{evaluations} presents the evaluation results for different RL agents and discusses the findings we seek to establish. Section~\ref{conclusion} concludes the whole paper.

%% file: sections/Related_Work.tex
\section{Related Work}\label{related work}

Reinforcement Learning (RL) serves as a powerful framework
for building automated quantitative trading systems,
enabling agents to explore complex market environments
and continually update their policies to optimize strategies
and maximize returns.
\citet{neuneier1995optimal} was the first to introduce deep reinforcement learning into the financial domain, represented by the use of Q-learning. Subsequently, \citet{moody1998performance,Moody2001Learning} applied actor-based RL to finance and employed recurrent neural networks (RNNs).
Since then, a wide variety of RL methods have been explored.
These studies usually rely not only on historical price and volume data, but also on technical indicators and/or other handcrafted factors.
For example, \citet{Zhang2019Deep} compared algorithms such as Deep Q-learning Networks (DQN), Policy Gradients (PG) and Advantage Actor-Critic (A2C) incorporated both discrete and continuous action spaces, and carefully designed the reward function.
\citet{Zou2024Deep} adopted a combination of an LSTM network and the PPO algorithm, while \citet{Huang2024A} proposed a novel BiLSTM-Attention architecture coupled with the RL SARSA algorithm.
All of these models take technical indicators or factors as part of their inputs.

There are also some attempts in RL that do not rely on handcrafted features, but instead start directly from raw price and volume data.
\citet{Deng2016Deep} used the deep neural network to extract features, 
which were then input into an RL agent with RNN.
\citet{Liang2018Adversarial, Theate2021Deep} input price and volume data into RL agents and applied different algorithms for RL.
\citet{Taghian2021Reinforcement} employed an encoder-decoder architecture
to extract features from raw data, which were then fed into the RL agent.

Financial news can significantly influence market price movements and should be considered as an additional state input, distinct from price information.
\citet{bollen2011twitter} was the first to focus on the impact of news on financial price movements, analyzing text sentiment using OpinionFinder and the Google Profile of Mood States (GPOMS).
\citet{loughran2011liability} analyzed financial text sentiment using a dictionary-based approach.
Although these methods enable large-scale processing of text sentiment, they still exhibit significant manual involvement.
In recent years, the development of large language models (LLMs) and their strong performance in Natural Language Processing (NLP) have provided an alternative approach for financial text sentiment analysis\cite{huang2023finbert,wu2023bloomberggpt}.
\citet{Liu2020FinRL,Liu2022FinrlMeta} incorporated the sentiment scores assigned by LLM to news together with price data as the state input to the reinforcement learning agent. \citet{Unnikrishnan2024Financial} constructed a sentiment-based reward for integration of sentiment analysis.
\citet{Benhenda2025FinRL-DeepSeek} applied sentiment scores to perturbatively adjust the agent’s decision-making actions.
\citet{Arshad2025FinRL} manually aligned the sentiment scores with price data using a market-aware module before inputting them into the agent.
These studies incorporate news information into reinforcement learning through manually specified rules and/or also include technical indicators.
In our work, we explore whether news information can be directly incorporated into a reinforcement learning framework without handcrafted features or manually designed rules.

%% file: sections/Methodology_v3.tex
\section{Methodology}
\label{methodology}

The core process of framework involves:
(1) leveraging LLM to extract sentiment and risk signals from financial news and integrating them with historical market prices into a time-series input;
(2) processing the integrated sequence through a LSTM or Transformer-based network within the RL agent to learn temporal patterns;
and (3) generating trading actions via a policy head optimized with RL to maximize financial returns;
(4) tuning model hyperparameters using the validation set and evaluating model via backtesting on the test set. In the following, we will present the detailed  architecture of our proposed framework.

\subsection{Sentiment Features Extraction}
\begin{figure*}[t!]
  \vspace{-\intextsep}
  \centering
  \includegraphics[width=0.9\linewidth]{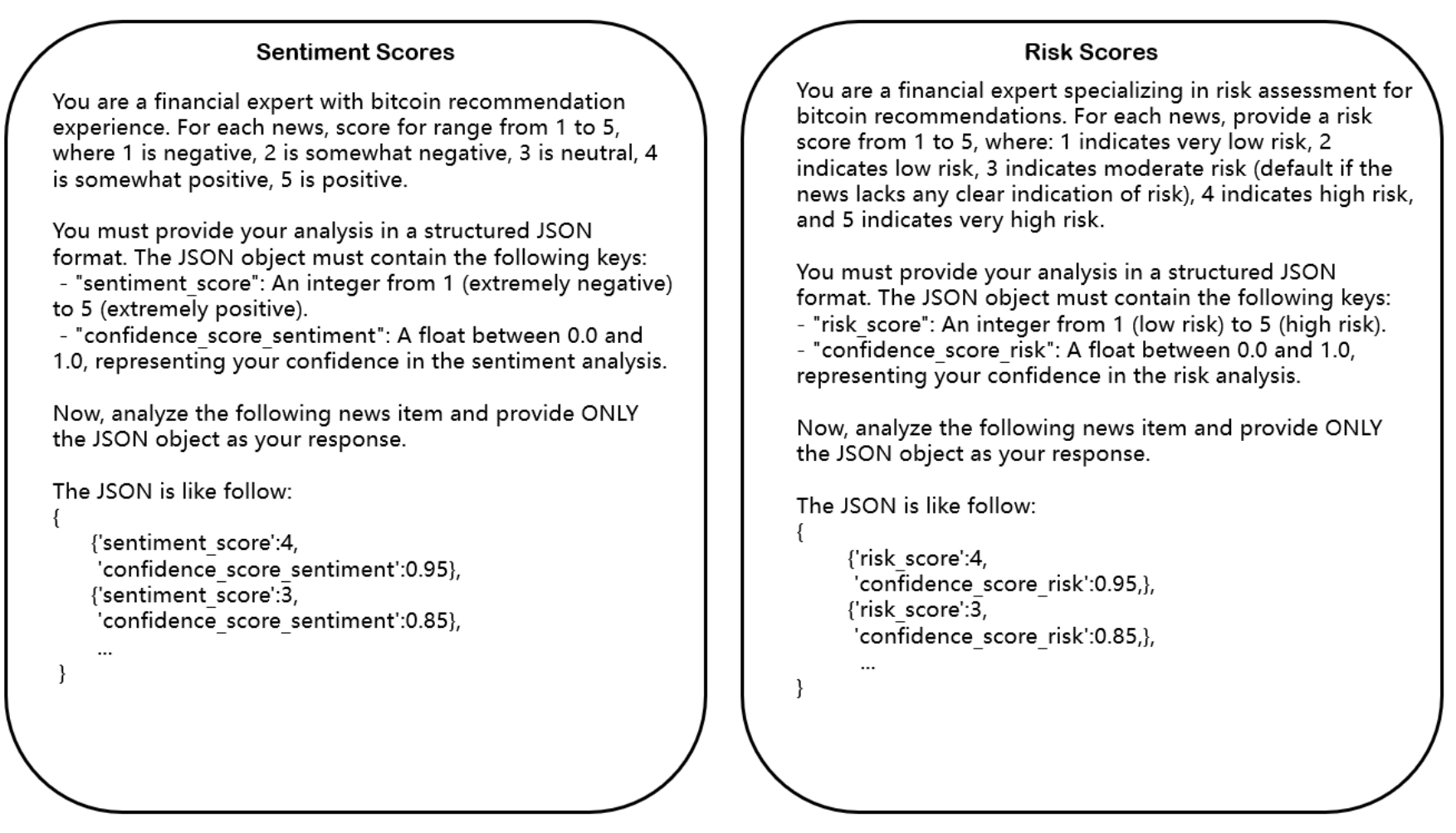}
  \caption{The format of input prompt that guides LLMs to generate sentiment scores and risk scores. It consists the task specification, the scoring system and the output format example.
}
  \label{fig:prompt_design}
  \vspace{-\intextsep}
\end{figure*}

The first stage of the framework involves extracting sentiment scores and risk scores from
financial news using LLMs with a robust and informative prompt following \cite{dong2024fnspid,Benhenda2025FinRL}
The prompt consists of a task specification that defines the analysis objective and a integer scoring system (i.e., 1–5 for sentiment and risk levels).
The template of the prompt is illustrated in Figure \ref{fig:prompt_design}.
We input the prompt into the \texttt{Gemini-2.5-flash} model 
\cite{comanici2025gemini} to assign sentiment and risk scores to the news items.
To maximize the utilization of API resources, the news items are processed in batches for scoring.
We have validated that, as long as the context length limit is not exceeded, the impact on the scoring results is negligible compared to the inherent randomness of the LLM itself.

\subsection{Reinforcement Learning Achitectures}
We consider both on-policy and off-policy reinforcement learning algorithms.
For off-policy learning, we employ Double DQN (DDQN), which is one of the enhancements of Deep Q-Network  algorithm.
For on-policy learning, we use a variant algorithm of the Proximal Policy Optimization (PPO), the Group Relative Policy Optimization (GRPO) algorithm.

Traditional Q-learning maintains a Q-value table and iteratively updates the Q-value through the Bellman equation. This kind of algorithm will have the problem of dimensionality disaster when the state/action space becomes extremely large.
DQN combines Q-learning methods with deep neural networks (DNN) to estimate Q-value, calculates the temporal-difference loss (TD-loss), and perform a gradient descent step to make an update. Among the enhancements of DQN, DDQN mitigates the overestimation bias of Q-values by decoupling action selection from value estimation. Specifically, it uses the current Q-network to select actions while employing the target network to evaluate the Q-values. 
The TD-loss used in DDQN is:
\begin{equation}
    \textit{L}(\theta)=\mathbb{E}[(r+\gamma \mathop{\max}\limits_{a'}\hat{Q}(s',\mathop{\max}\limits_{a}Q(s',a))-Q(s,a))^2]
\end{equation}
where $\hat{Q}$ is the target network with weights parameter $\theta$, $s'$ denotes the next-time-step state, and $a'$denotes the action that maximizes $\hat{Q}$. 

PPO is an RL algorithm introduced by OpenAI and widely used in financial trading tasks~\cite{Schulman2017Proximal, Lele2020Stock}, it builds upon the principles of Trust Region Policy Optimization (TRPO).
PPO simplifies TRPO by replacing the constraint with a specialized clipped objective function. This function restricts the ratio between the probabilities of the new and old policies, preventing the policy from too rapid changes that could destabilize training. Specifically, the optimizing objective is:
\begin{equation}\label{PPO_objective}
     \textit{L}(\pi_{\theta})=\mathbb{E}\left[\min(r_{t}(\theta)A_{t},\mathop{\mathrm{clip}}(r_{t}(\theta), 1-\epsilon, 1+\epsilon)A_{t})\right]
\end{equation}
where $r_{t}(\theta)=\pi_{\theta}(a_t|s_t) / \pi_{\theta,\, old}(a_t|s_t)$ denotes the probability ratio between new policy $\pi_{\theta}$ and old policy $\pi_{\theta,\, old}$, $A_{t}$ is the advantage function at time $t$, $\epsilon$ is the clipping parameter that restricts the large changes between old and new policies.

GRPO is an improved version of PPO, proposed by the DeepSeek team~\cite{shao2024deepseekmath}. In GRPO algorithm, $A_{t}$ is replaced with $\hat{A}_{i,t}$ in Eq.(\ref{PPO_objective}), where $\hat{A}_{i,t}=\frac{r_i-\mathrm{mean}(\boldsymbol{r})}{\mathrm{std}(\boldsymbol{r})}$ represents the relative reward between different groups without the use of critic network. Since the value function used in PPO is usually implemented as a separate model of comparable size to the policy network, it imposes a significant memory and computational overhead.

We investigate two classes of network architectures in our proposed method: multilayer perceptron (MLP) and sequence-based architectures, specifically LSTM and Transformer networks ~\cite{Graves2012Long, Vaswani2017Attention}.
We use MLPs to assess the impact of temporal sequences in our framework, in which data at a single time point is utilized.
In sequence-based nets cases, we first feed the state into the sequence network, then input the output from the last time step into a single-layer MLP to obtain the action.
Specifically, In the case of Transformer, we adopt an encoder stack structure with learnable positional encodings, a linear input projection maps input features to the model dimension before the encoder.
It should be noted that 
when using different network architectures, the same MLP/sequence-based network is consistently applied across all components within a given algorithm (DDQN/GRPO).
The AdamW optimizer~\cite{Loshchilov2017Decoupled} is adopted for weight optimization throughout the training process.

\subsection{Dataset Preprocessing}
\begin{figure}[t!]
  \centering
  \includegraphics[width=1.0\columnwidth]{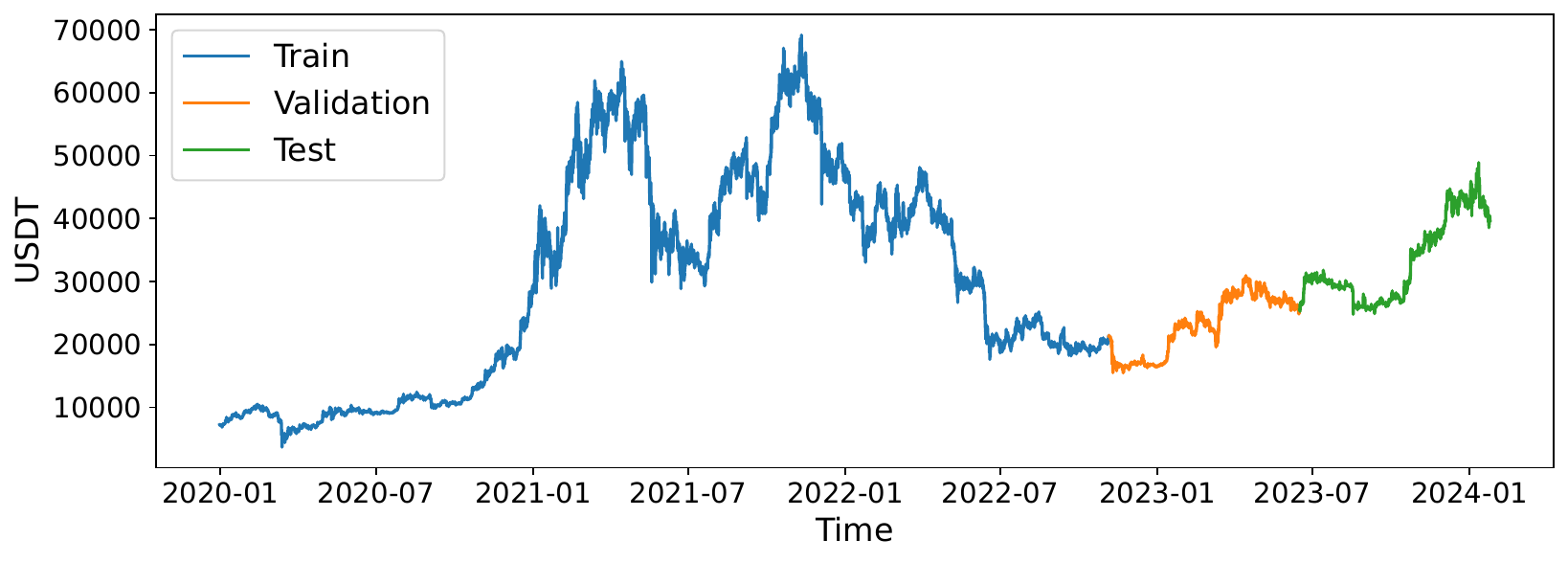}
  \caption{BTC (1-minute) Prices timeline division in our framework. Blue part: training set. Orange part: validation set. Green part: test set.
}
  \label{fig:BTC_price}
\end{figure}

The whole dataset, including market 1-minute OHLCV (i.e. open, high, low, close, volume)
time-series sourced from Binance Exchange\footnote{https://data.binance.vision/} for BTC/USDT and news text scrapped from Yahoo Finance related to Bitcoin\footnote{\url{https://huggingface.co/datasets/edaschau/bitcoin_news}}.
We set the sentiment and risk scores for the interval between two successive news items to be governed by the preceding one.
The time range is from 2019-12-31 00:00:00 to 2024-01-24 21:48:00.
We divided the whole dataset
into training, validation and test area using a chronological split to prevent look-ahead bias and ensure realistic performance evaluation, see Figure~\ref{fig:BTC_price}.
The training set encompasses the initial 70\% of the timeline, the validation set covers the subsequent segment from 70\% to 85\%, and the test set comprises the remaining portion from 85\% to the end.
During training, the agent learns from the training interval, while the validation set guides hyper-parameter tuning and agent selection, final performance is assessed on the test set.

\subsection{Hyper-parameters Tuning}\label{Hyper-parameters Tuning}
It is well-known that reinforcement learning is sensitive to the choice of hyper-parameters.
To make a fair comparison, we tune the hyper-parameters
to efficiently explore the hyper-parameter space across all model configurations.
Our tuning process is designed as follows: agents are trained on the training set, while performance on the validation set guides both hyper-parameter selection and early stopping.
During training, for each hyper-parameter combination (i.e. trial), if the highest average return on the validation set does not increase for five continuous evaluations, the trial is terminated prematurely.
These hyper-parameters including network architecture hyper-parameters (e.g. hidden dimensions, number of layers), sequential model hyper-parameters (e.g. window size, number of attention heads), agents training hyper-parameters (e.g., learning rate, batch size), and algorithm-specific hyper-parameters (e.g. exploration rate for DDQN, clip ratio for GRPO).
The detailed hyper-parameters and their corresponding value ranges are summarized in Appendix~\ref{Supplementary for hyper-parameters tuning} Table~\ref{tab:hyperparameters}.

%% file: sections/Experiment.tex
\section{Performance Evaluations}\label{evaluations}
As a proof-of-concept study, we consider a simple discrete action space with only three actions: short, long one BTC, and hold.
We consider a stop-loss/take-profit threshold of 0.1\% to simulate practical risk management constraints.
We randomly sample 3,000-minute consecutive trading periods from the dataset for training and validation.
Specifically, during hyperparameter tuning, we use the statistical mean of cumulative returns over 256 periods sampled from the validation set as the optimization target.

\begin{figure*}[t!]
  \centering
  \begin{subfigure}{0.49\textwidth}
    \centering
    \includegraphics[width=\linewidth]{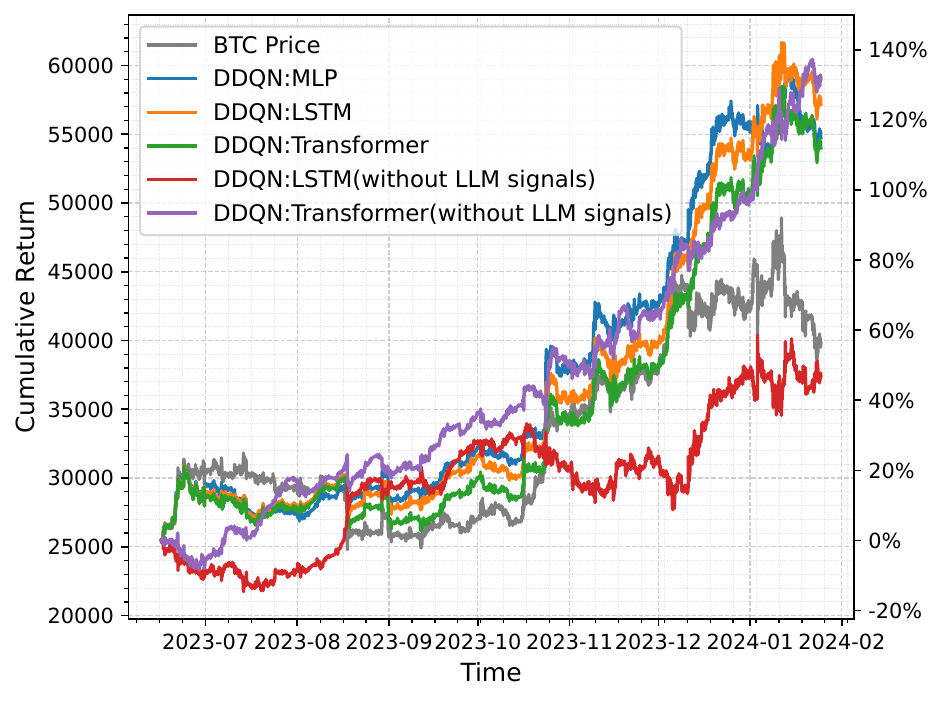}
    \subcaption{DDQN Top1}
  \end{subfigure}\hfill
  \begin{subfigure}{0.49\textwidth}
    \centering
    \includegraphics[width=\linewidth]{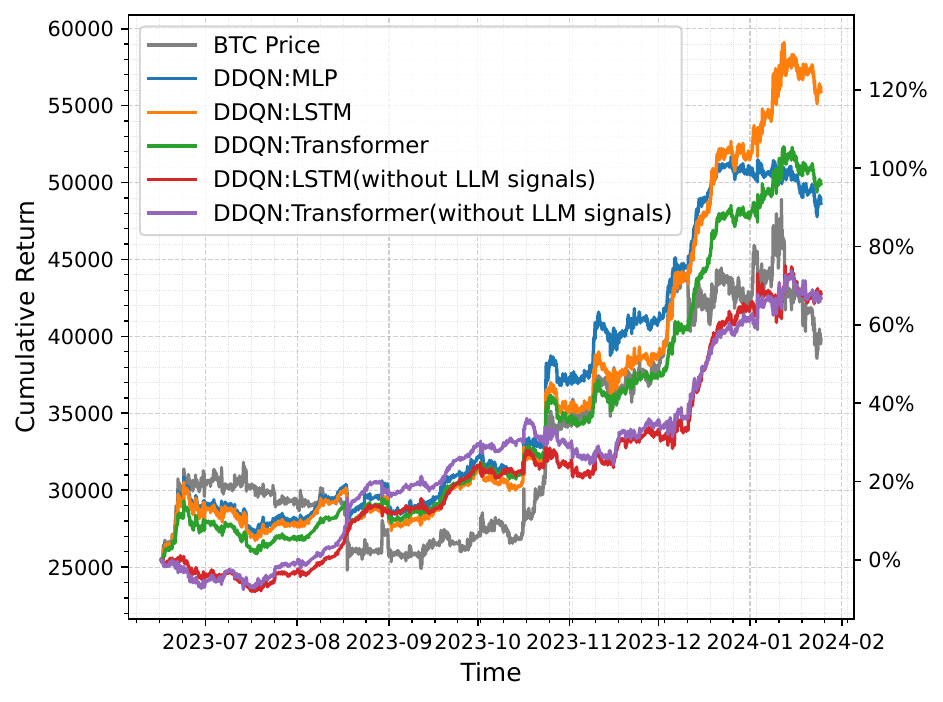}
    \subcaption{DDQN Top10}
  \end{subfigure}

  \begin{subfigure}{0.49\textwidth}
    \centering
    \includegraphics[width=\linewidth]{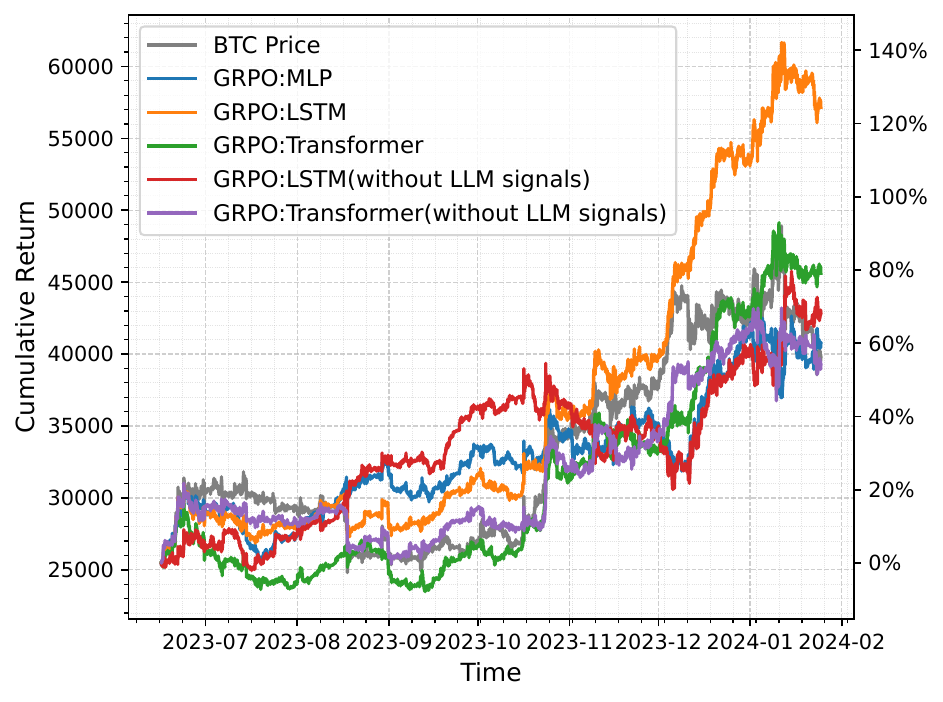}
    \subcaption{GRPO Top1}
  \end{subfigure}\hfill
  \begin{subfigure}{0.49\textwidth}
    \centering
    \includegraphics[width=\linewidth]{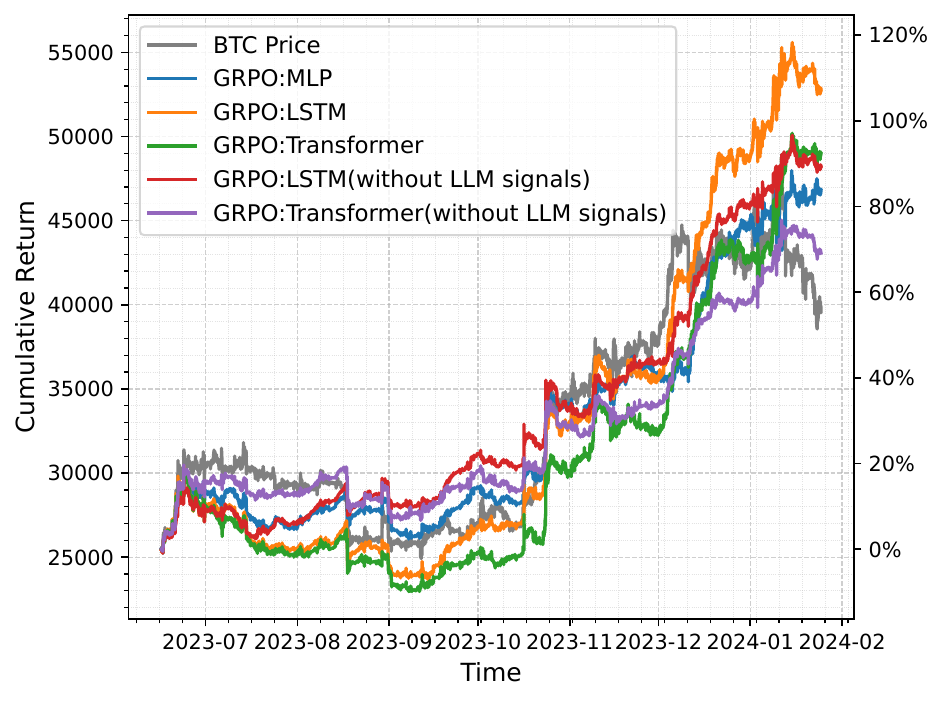}
    \subcaption{GRPO Top10}
  \end{subfigure}

  \caption{Generalization performance of validation-selected Top-K agents on the BTC test set: DDQN (top row) and GRPO (bottom row). Each panel marks cumulative return on the test set in USDT terms (left-side axis) and percentage terms (right-side axis). Columns correspond to $K\in\{1,10\}$; for $K=10$, curves are averaged across the top-$K$ agents ranked by validation performance. The gray curve denotes the BTC prices baseline, while colored curves indicate distinct network architectures (MLP, LSTM, Transformer) and the presence/absence of an LLM-derived news-sentiment signals. }
  \label{fig:backtest}
\end{figure*}

Finally, we similarly sample 256 periods from the test set to compute the average cumulative return over 3,000-minute trading periods.
Our evaluation results are summarized in Table~\ref{tab:results} as following, 
For each algorithm (DDQN, GRPO), we test different network architectures (MLP, LSTM, and Transformer).
In addition to the full models that leverage LLM-derived news–sentiment signals, we conduct ablations for the LSTM and Transformer backbones in which the LLM signals are disabled and only market time-series inputs are provided, which isolates the contribution of news sentiment to trading performance.
We consider two simple ways of utilizing the agents. The first is to use the agent that performs best on the validation set. The second is to use the top 10 agents with the best performance on the validation set and take the statistical average of their results to reduce the impact of randomness.
In addition to evaluating the average results over 3,000-minute periods, we also perform a full backtest over the entire test period, as show in Figure~\ref{fig:backtest}, the cumulative returns are shown in Table~\ref{tab:backtest_returns}, where the BTC market baseline return is 56\% over the test period.

\begin{table*}[!ht]
\centering
\caption{Averaged cumulative returns (in USDT) for different RL algorithms with various network architectures, the optimal performance of the agents are highlighted in bold font.}
\label{tab:results}
\setlength{\tabcolsep}{6pt}
\renewcommand{\arraystretch}{1.1}

\begin{tabular}{l *{2}{cc}} 
\toprule
& \multicolumn{2}{c}{DDQN} & \multicolumn{2}{c}{GRPO} \\
\cmidrule(lr){2-3}\cmidrule(lr){4-5}
Networks & Top1 & Top10 & Top1 & Top10 \\
\midrule
MLP         &  80.6 & 153   & 203.2 & 151.5 \\
LSTM        & \textbf{329.8} & \textbf{338} & \textbf{447.5} & \textbf{289.5} \\
Transformer & 307.1 & 223.8 & 227     & 219.4    \\
LSTM (Without LLM signal) & 201.9 & 118.1 & 135.4  & 265.9 \\
Transformer (Without LLM signal) & 283.8 & 199.3 & 272.1 & 224.9 \\
\bottomrule
\end{tabular}
\end{table*}

\begin{table*}[!htbp]
\centering
\caption{Full backtest cumulative returns (percentage change) for different RL algorithms with various network architectures, the optimal performance of the agents are highlighted in bold font.}
\label{tab:backtest_returns}
\setlength{\tabcolsep}{6pt}
\renewcommand{\arraystretch}{1.1}

\begin{tabular}{l *{2}{cc}} 
\toprule
& \multicolumn{2}{c}{DDQN} & \multicolumn{2}{c}{GRPO} \\
\cmidrule(lr){2-3}\cmidrule(lr){4-5}
Networks & Top1 & Top10 & Top1 & Top10 \\
\midrule
MLP         &  114.9\%      & 91 \%             & 59.9\% & 83.7\% \\
LSTM        & 124.5\%       & \textbf{119\%}    & \textbf{124.5\%} & \textbf{106.8\%} \\
Transformer & 112\%         & 95.8\%            & 79.1\%     & 92.2\%    \\
LSTM (Without LLM signal) & 47\%      & 67.8\%        & 68.3\%  & 89\% \\
Transformer (Without LLM signal) & \textbf{131.8\%}         & 66.7\% & 54.4\% & 69.2\% \\
\bottomrule
\end{tabular}
\end{table*}

Our evaluation results shows that the proposed news-aware RL framework achieves higher cumulative returns than the BTC market benchmarks on the test set. This advantage holds for both DDQN and GRPO algorithms, indicating the robustness of our framework to model choice variance.

On the other hand, results in Figure~\ref{fig:backtest}, Table~\ref{tab:results} and Table~\ref{tab:backtest_returns} show the architectural contrast. For LSTM-based RL agents, adding LLM-derived news sentiment consistently raises cumulative returns relative to the LSTM without news, indicating the contribution of news information. For Transformer-based agents , the contribution from news is weaker.
This suggests that Transformer encoder may be less sensitive than LSTM at dealing with news information and capturing temporal dependencies between time-series data. 

Moreover, sequence-based agents (LSTM, Transformer) consistently outperform the MLP-based agents for both DDQN and GRPO algorithms, confirming the value of modeling prices and sentiment signals as continuous time-series rather than isolated inputs. Between the two sequence models, LSTM-based agents outperform Transformer-based agents in our setting. One potential reason is that Transformers we used are not explicitly tailored for time-series modeling and we did not introduce time-series specific adaptations to the Transformer architecture in our framework, whereas LSTM can naturally process time-series data with inherent causal structures.

Overall, the aforementioned evalution results illustrate that our proposed framework can exceed the BTC market benchmarks.
Incorporating LLM-derived news-sentiment inputs yields higher returns than the corresponding model without LLM signals, confirming the value of news analysis. Moreover, sequence-based agents (LSTM and Transformer) consistently outperform MLP agents, indicating that modeling prices and sentiment as continuous sequences rather than isolated inputs is essential for effective algorithmic trading.

%% file: sections/Conclusion.tex
\section{Conclusion}\label{conclusion}
This paper introduces a news-aware RL framework for financial trading that leverages sequence-based networks to process raw market prices and volume with news sentiment features directly.
We show that even without handcrafted features or manually designed rules,
RL agents can feasibly utilize news sentiment features derived from large language models, together with price data, as time-series inputs processed by LSTM or Transformer architectures.
Evaluation results show that our proposed framework can outperform the market baseline in both DDQN/GRPO algorithms, they also demonstrate the importance of news events utilization and the effectiveness of leveraging sequence-based network architectures to capture temporal dependencies between time-series data.
It opens a promising direction for future research on incorporating news information into financial market trading with minimal or no reliance on manual intervention.

This work serves as a proof of concept, focusing on evaluating the feasibility of incorporating LLM-derived news sentiment signals and sequence-based network architectures without handcrafted features.
Further research can develop fully optimized trading strategies for practical deployment and return maximization.
For instance, rather than the simple averaging of top-performing agents adopted in our study for evaluation, the more efficient utilization of hierarchical multi-agent frameworks probably be required for enhanced collective decision-making in the future.
Furthermore, the action space and risk management employed in this study are simplified, and practical trading applications would require more advanced designs.
In addition, future work could explore architectures better suited to time-series modeling to capture market prices and sentiment dynamics more effectively.

%% file: sections/Appendix.tex
\onecolumn
\section{Supplementary for hyper-parameters tuning}\label{Supplementary for hyper-parameters tuning}
This appendix section provides the supplementary materials for subsection~\ref{Hyper-parameters Tuning} in Section~\ref{methodology}. In this study, we tune the hyper-parameters using \texttt{optuna} with TPE (Tree-structured Parzen Estimator) algorithm~\cite{Bergstra2011Algorithms}. Table~\ref{tab:hyperparameters} shows the tuned hyper-parameters for RL structure with different algorithms and network architectures and their corresponding value ranges that we used in this paper.
\begin{table*}[htbp]
\centering
\caption{Hyper-parameters and tuning ranges/values.}
\label{tab:hyperparameters}
\begin{adjustbox}{width=\linewidth}
\begin{tabular}{p{0.55\linewidth} p{0.4\linewidth}}
\toprule
\textbf{Description} & \textbf{Range/Values} \\
\midrule
Window size for sequence models & 10 to 50 (log scale) \\
Hidden dimension size for LSTM & [32, 64, 128] \\
Number of layers for LSTM/Transformer & [1, 2] (LSTM), 1 to 3 (Transformer) \\
Position encoder standard deviation in Transformer & 0.02 to 1 (log scale)  \\
Number of attention heads in Transformer & [2, 4] \\
Feedforward dimension in Transformer & [32, 64, 128] \\
First hidden layer size in MLP & [32, 64, 128] \\
Second hidden layer size in MLP & [32, 64, 128] \\
Discount factor for future rewards & 0.90 to 0.995 (log scale) \\
Gradient clipping norm & 0.1 to 4.0 (log scale) \\
State value target update rate & [0, 0.01] \\
Learning rate for optimizer & 2e-6 to 1e-3 (log scale) \\
Weight decay coefficient & 1e-5 to 1e-2 (log scale) \\
Training batch size & [32, 128, 512] \\
Policy update repetition times & [1, 2] (off-policy), [4, 8] (on-policy) \\
Horizon length for training & max\_step × [2, 4, 8] \\
Replay buffer size & horizon\_len × [2, 4, 8] \\
Exploration rate for DDQN & 0.005 to 0.125 (log scale) \\
Epsilon decay rate for DDQN & [0.99995, 0.99999, 0.999999] \\
Soft update rate for target network & 1e-3 to 1e-2 (log scale) \\
GAE parameter for PPO & 0.9 to 0.99  \\
Clipping ratio for PPO & 0.1 to 0.2  \\
Target KL divergence for PPO & 0.005 to 0.02 (log scale) \\
Entropy coefficient for PPO & 0.001 to 0.1 (log scale) \\
Value function coefficient for PPO & 0.1 to 1.0 (log scale) \\
\bottomrule
\end{tabular}
\end{adjustbox}
\end{table*}